\documentclass{article}
\usepackage[T1]{fontenc}
\UseRawInputEncoding
\usepackage[english]{babel}
\usepackage{bm}
\usepackage{slashed}
\usepackage{amsmath}
\usepackage{amsthm}
\usepackage{amssymb}
\usepackage{xcolor}

\makeatletter

\newcommand*\LyXZeroWidthSpace{\hspace{0pt}}

\numberwithin{equation}{section}
\numberwithin{figure}{section}
\newcommand{\lyxaddress}[1]{
	\par {\raggedright #1
	\vspace{1.4em}
	\noindent\par}
}
\theoremstyle{plain}
\newtheorem*{prop*}{\protect\propositionname}
\theoremstyle{definition}
\newtheorem*{problem*}{\protect\problemname}

\@ifundefined{date}{}{\date{}}
\usepackage{titling}
\pretitle{%
  \begin{flushleft}
  OUJ-FTC-17\\[1em]
  \LARGE
}
\posttitle{\end{flushleft}}

\makeatother

\addto\captionsenglish{\renewcommand{\problemname}{Problem}}
\addto\captionsenglish{\renewcommand{\propositionname}{Proposition}}
\providecommand{\problemname}{Problem}
\providecommand{\propositionname}{Proposition}

\begin{document}
\title{A Constructive Definition of Space via Dynamical Evolution and Observational
Acts}

\author{So Katagiri\thanks{So.Katagiri@gmail.com}}

\maketitle

\lyxaddress{\textit{Major in Complex Systems Science, Graduate School of Science and Engineering, Ibaraki University, Nakanarusawa-cho, Hitachi, 316-8511, Japan.}}
\begin{abstract}
We propose a constructive and dynamical redefinition of spatial structure,
grounded in the interplay between mechanical evolution and observational
acts. Rather than presupposing space as a static background, we interpret
space as an emergent entity that arises through observational acts.
\color{black}Given a time evolution and a physically specified measurement context, we construct a reachability structure and a context-dependent measurable structure.
\color{black}
Using the framework of pre-topologies, measurable structures, and
the GNS construction, we analyze how the choice of observables and
the system's time evolution dynamically determine the topological
and measure-theoretic features of space.
\color{black}Time evolution and the measurement context are inputs to this construction. Deriving time itself or a complete dynamics for the selection and stabilization of the measurement context lies outside the scope of this paper.
\color{black}
 This approach highlights
the observer-dependent and context-sensitive nature of spatial concepts
in both classical and quantum domains.
\end{abstract}

\section{Introduction}

In mathematics, the concept of ``space'' appears in a wide variety
of forms---topological spaces, metric spaces, measure spaces, linear
spaces, Hilbert spaces, and more. Despite their diversity, these structures
are all called ``spaces'', often without a clearly articulated reason
for their common designation. While the mathematical operations applied
within these spaces are rigorously defined, the notion of space itself
is often left to intuitive understanding, heavily dependent on context.
Thus, the term ``space'' functions not so much as a rigorously defined
mathematical object, but rather as a conceptual apparatus---one that
resides outside formal mathematics and is anchored in the intuition
and contextual judgment of the mathematician.

This conceptual ambiguity not only signals a lack of foundational
clarity, but also allows implicit philosophical assumptions to infiltrate
mathematical reasoning---often without scrutiny.

Even in rigorous disciplines such as functional analysis or algebraic
geometry---fields that emphasize internal consistency---the notion
of ``space'' often functions more as an intuitive scaffold than as
a strictly defined object.

In light of these issues, we ask: What is space---not as a physical
or metaphysical object, but as a mathematical construct shaped by
the observer's methodology and purpose? The mathematical concept of
space, we argue, is deeply rooted in the mathematician's observational
acts and constructive motives. It is perhaps more accurate to see
it as a reflection of the mathematician's spatial intuition than as
a strictly defined object.

In what follows, we approach this question constructively: we reinterpret
spatial structures not as static entities, but as dynamical constructs
shaped by time evolution and the act of observation.

By exploring this question, we aim to shed light on the internal structure
of the spatial concept in mathematics and reveal that even abstract
mathematical spaces are often constrained within the classical framework
of space that we intuitively understand. This line of thought aligns
with Kato's Leibnizian relational view of space\cite{key-1}, as well
as modern topos-theoretic perspectives, in which space is not a pre-existing
container but a structure that emerges as a result of constructive
activity.

Mathematics is an indispensable tool for physics. Since Newton, and
through the development of relativity and quantum theory, physics
has evolved beyond the classical intuition of space and time. Meanwhile,
mathematics---particularly through foundational studies originating
from problems like the heat equation---has taken a different path.
Although the two disciplines occasionally interact in contexts such
as gauge theory or general relativity, they largely follow separate
trajectories. In this context, it is not evident that modern mathematics
functions as an appropriate tool for modern physics.

For instance, quantum field theory and path integral formulations
still lack fully rigorous mathematical definitions. Moreover, when
fundamental questions regarding the nature of spacetime are raised,
traditional spatial concepts inherited from classical physics may
no longer be valid---particularly in light of Bohr's complementarity
and the breakdown of classical realism in quantum contexts.

With this in mind, the present paper proposes a constructive redefinition
of space that is grounded in physical dynamics and observational acts.
We begin with a reinterpretation of topological concepts from the
perspective of dynamical systems, introducing the notion of reachability
domains and pre-topologies. We then consider measurable structures
as projections determined by observation, and finally, we examine
the role of non-commutative observables in quantum theory. Through
these investigations, we show how spatial structures arise dynamically
and contextually, depending on the system's evolution and the observer's
choices.

This perspective bears resemblance to several prior approaches, yet
differs in crucial ways that highlight its novelty.

A related line of inquiry can be found in Connes' noncommutative geometry
\cite{key-7}, which extends the Gel'fand--Naimark duality---originally
linking commutative $C^{*}$-algebras and topological spaces---to
the noncommutative setting. In Connes' framework, geometry is encoded
algebraically via $C^{*}$-algebras, leading to the notion of ``quantum
spaces.'' 
In contrast, we argue that it is precisely the time evolution of a
system and the observer's choice of observable that jointly give rise
to spatial structures. The topology of space is dynamically determined
by the system's evolution, while the measure structure emerges from
the act of observation. In quantum theory, this leads naturally to
the context-dependence of mutually incompatible observables---i.e.,
the impossibility of simultaneous measurement. 
\color{black}The thermal time hypothesis of Connes and Rovelli addresses a different question: in the present paper the time evolution is taken as an input.\color{black}

The groundbreaking aspect of Connes'approach lies in the observation that, once the algebra of ``functions'' on space becomes noncommutative, the very notion of a point loses its conventional meaning. In this sense, noncommutative geometry deals with what is often called a point-less geometry. By contrast, in our constructive framework, the set of observables under consideration is always taken to be a mutually commuting set within a chosen context, so that within this restricted set the assignment of values is still meaningful. In this sense, the notion of a ``point'' retains a partial operational meaning.

\color{black}
Connes' noncommutative geometry describes geometric structures using algebraic data. This paper focuses on making explicit the roles of reachability under a specified time evolution and distinctions realized within a physically specified measurement context in constructing space. A complete dynamical account of how the measurement context itself is selected and stabilized is beyond the scope of this paper.
\color{black}

\color{black}
In Rovelli's relational quantum mechanics, physical facts are relative to interactions between physical systems, and an observer is itself treated as a physical system \cite{key-9,key-16}. Measurement in classical mechanics has also been examined in terms of the interaction between a system and an apparatus and their relative descriptions, drawing on Everett's ``relative state'' formulation \cite{key-20,key-18}. Against this background of physically constrained observation, this paper focuses on the measurable structures defined by distinctions realized within a specified measurement context.
\color{black}

As another line of discussion, following Born's proposal of the reciprocity principle\cite{Born-1}, there has been recent interest in quantum gravity scenarios that take seriously the possibility that momentum space itself may be curved\cite{Boyle-1}\cite{Glikman-1}. Such approaches often regard phase space, rather than spacetime, as the fundamental arena, with the dynamical nature of the observer influencing the reconstruction of space\cite{Magid-1}. This perspective shares common ground with our view that space should be constructed from dynamics and observation. However, from our standpoint, momentum is regarded merely as one possible observable, and in this sense our formulation may be considered more general. \footnote{An earlier version of this work was originally published in Japanese as \cite{key-2}.}

\color{black}By treating a commutative subalgebra as a measurement context, this paper has a point of contact with quantum contextuality and topos-theoretic approaches \cite{key-13}. Nevertheless, facts relative to relations between physical systems, contexts comprising mutually compatible observables, and consistency between different contexts are distinct questions. The present paper considers the construction of a measurable structure within a specified context.\color{black}

\section{Pre-Topological Spaces and Reachability Domains: A Theory of Space
Without Observers}

In this section, we revisit the concept of topological space from
the standpoint of pre-topology, using the notion of reachability domains
derived from the time evolution of dynamical systems. We will demonstrate
how such reachability structures naturally induce a pre-topological
structure on a set. Subsequently, by analyzing these reachability
domains, we provide a dynamical interpretation of topological spaces
themselves. From this perspective, a topological space can be understood
as \color{black}a stabilized reachability structure within a dynamical system.\color{black} \color{black} From this perspective, a topological structure arises when the reachability operator is idempotent, so that reapplying it does not enlarge the reachability domain.\color{black}

\subsection{Dynamical Systems}

A dynamical system is a theoretical framework for describing physical
quantities that evolve over time. Let $x(t)$ denote the state of
a system at time $t$. If we denote the initial state at time $t=0$
as $x_{0}$, then the time evolution of the system after $t$ units
of time can be represented as
\begin{equation}
x(t)=U(x_{0},t).
\end{equation}

Here, $U$ is a function representing the dynamical evolution. Although
this notation may resemble the unitary evolution operator in quantum
theory, it should be noted that $U$ is not assumed to be unitary
in general. That is, we are not restricting ourselves to Hamiltonian
systems\footnote{Note that the only primitives assumed here are a set and its dynamics; no metric or pre-existing topological structure is presupposed}.

Importantly, the term ``dynamical system'' is used here in a broad
sense---it refers not only to Hamiltonian mechanics, but to any system
that exhibits time-dependent evolution.

By definition, the evolution function satisfies:
\begin{equation}
U(x_{0},0)=x_{0},
\end{equation}
which simply asserts that the state remains unchanged at time zero.

The trajectory of a point $x(t)$ under such a system, given an initial
state $x_{0}$, a dynamical evolution $U$, and a time horizon $T$,
is defined by the set
\begin{equation}
D_U(x_{0},T)=\{U(x_0,t)|0\leq t\leq T\}.
\end{equation}
This trajectory depends on the initial condition, the dynamics of
the system, and the duration of evolution. If the system evolves in
discrete time, then $T\in\mathbb{Z}$, whereas for continuous systems,
we take $T\in\mathbb{R}$.

This immediately raises the question: what kinds of structures can
serve as ``time''? At the very least, it is conceivable that time
could be modeled not only by the real or integer numbers but also
by more general algebraic structures such as monoids, groups, or partially
ordered sets (posets).

Although one could consider more exotic scenarios---such as systems
with multiple time parameters or higher-dimensional time---we restrict
ourselves here to the simplest case: a single, linear notion of time\footnote{ Generalization to the case where time is modeled as a poset is technically straightforward, and would naturally connect to Lieb-Robinson-type causal structures in many-body quantum systems, suggesting further links to Lieb's thermodynamic formulation\cite{Lieb-1}. A detailed treatment is beyond the scope of the present work.}.

When time is modeled as a monoid, repeated composition of time evolutions
becomes well-defined and naturally interpretable. In such a case,
time admits a binary operation (typically denoted by composition or
addition) and a unit element, enabling the successive application
of the evolution operator.

Let us consider a dynamical evolution operator 
\begin{equation}
U(t):X\to X,
\end{equation}
where $x\in X$ evolves under $U$. If time is structured as a monoid,
then the following properties hold:
\begin{equation}
U(s)\circ U(t)=U(s\circ t),
\end{equation}
\begin{equation}
U(e)=id_{x},
\end{equation}
where $e$ is the identity element of the monoid, and $\text{id}_{x}$
is the identity map on $X$. In this case, since inverse elements
are not assumed to exist, the ``arrow of time'' is irreversible.

If time is modeled instead by a group, the existence of inverse elements
allows for time-reversal symmetry: for every $t$, there exists $-t$
such that $U(-t)$ is inverse of $U(t)$.

Alternatively, if time is represented by a partially ordered set (poset),
the time evolution respects a partial order structure. This interpretation
naturally encodes causal relationships---events evolve in a way that
preserves causal ordering.

This discussion can be summarized in the following slogans:
\begin{itemize}
\item Monoid: Linear repetition of time steps $t_{1},t_{2},t_{3}$
\item Group: Symmetry under time reversal $t\longleftrightarrow-t$
\item Poset: Branching and causal ordering through partial order 
\end{itemize}
In most physical theories, time is treated as a totally ordered set,
and we will adopt that assumption here as well. However, it is important
to note that in subsequent sections---especially when discussing
reachability domains, closure, and continuity---the nature of time
(whether totally ordered or structured as a general monoid) will influence
how these topological notions are defined and interpreted.

The nature of the space $X$, to which the evolving quantity $x(t)\in X$
belongs, can vary widely depending on the physical or mathematical
context. If $X$ is taken to be real space and the dynamical system
is governed by a Hamiltonian, we are in the setting of classical mechanics.
If $X$ represents a thermodynamic phase space and the dynamics follow
Onsager's theory, the framework corresponds to nonequilibrium thermodynamics.

More generally, $X$ may be an abstract set or a topological space.
In the context of quantum mechanics, for instance, $X$ can be interpreted
as a space of operators. Thus, the framework of dynamical systems
is flexible enough to encompass a wide variety of physical and mathematical
theories by appropriate choices of the underlying space $X$ and the
dynamics $U$.

\subsection{Reachability Domains}

We now introduce the concept of reachability, which plays a central
role in our constructive interpretation of space.

Given a dynamical system $U$ and a time horizon $T$, the reachability
domain of an initial region $X_{0}\subset X$ is defined as the union
of the trajectories emanating from all initial points in $X_{0}$
over the time interval $[0,T]$.

Formally, we define:
\color{black}
\begin{align}
\bar{X}_{0}^{T}(U)\equiv D_U(X_{0},T)\equiv\bigcup_{x_0\in X_0} D_U(x_{0},T)  \\
=\{U(x_{0},t)|x_{0}\in X_0,0\leq t\leq T\}= \mathrm{cl}_U^T(X_0).
\end{align}
\color{black}
Here, 
\begin{equation}
D_U(x_{0},T) := \{\,U(x_{0},t) \mid 0 \le t \le T\,\} .
\end{equation}
We use the overline notation $\bar{X}_{0}^{T}$ as a shorthand for $\mathrm{cl}_U^T(X_0)$, to suggest a closure
operation, anticipating the later identification of this object with
a kind of dynamical closure in the context of pre-topological structures.
Here $\mathrm{cl}_U^T$ plays the role of a closure operator; if it satisfies idempotency, the induced structure is a topology, otherwise a pre-topology.

Conceptually, this construction describes how a ``bundle of particles''
initially distributed over $X_{0}$ evolves over time under the dynamics
$U$. As we shall see, such reachability domains form the foundation
for inducing spatial structures dynamically---space emerges from
the collective evolution of matter or information.

\subsection{Example}

Let us now consider a simple illustrative example. Take the set

\begin{equation}
X=\{1,2,3,4,5\}.
\end{equation}
Define a dynamical system $U$ such that the state shifts one position
to the right at each unit time step:
\begin{equation}
U(x,1)=x+1.
\end{equation}
We impose periodic boundary conditions so that
\begin{equation}
U(5,1)=1.
\end{equation}
Let the initial region be
\begin{equation}
X_{0}=\{1,2,3\}.
\end{equation}
Then the reachability domain after one time step is:

\begin{equation}
\bar{X}_{0}^{T=1}(U)=\{1,2,3,4\}.
\end{equation}
This example illustrates how the structure of the reachability domain
evolves under the dynamics. A key observation is that when a subset
coincides with its reachability domain, it can be regarded as a closed
set in the pre-topological sense. In the next section, we develop
this interpretation more formally, defining closed and open sets within
a pre-topological framework based on dynamical behavior.

\subsection{Dynamical Interpretation of Pre-Topology}

A pre-topological space on a set $X$, denoted as $(\{X_{0}\subset X\},U,T)$,
is defined by equipping $X$with a dynamical system $U$ and a time
horizon $T$, and by examining how subsets $X_{0}\subset X$ evolve
over time.

Specifically, the structure of the pre-topology is determined by how
the reachability domains of initial regions $X_{0}$ behave under
the dynamics up to time $T$.

This approach can be seen as a dynamical concretization of the ideas
proposed by J$\ddot{\mathrm{u}}$rgen Jost \cite{key-5}, where spatial
structure is not imposed a priori but instead emerges from the behavior
of systems over time.

A key feature of this perspective is that the structure of the pre-topology
depends on the dynamics. By modifying the dynamical system $U$, the
pre-topological structure on $X$ also changes. Likewise, altering
the time horizon $T$ affects the reachability domains and thus reshapes
the structure of the pre-topological space.

In this way, we begin to see a direct link between dynamics and spatial
structure: space is no longer a fixed backdrop but a system-dependent,
emergent structure shaped by temporal evolution.

\subsection{Dynamical Interpretation of Closed Sets}

The notion of a closed set can be naturally reinterpreted through
a dynamical lens. In this context, a subset $X_{0}\subset X$ is considered
closed if, under the evolution of the dynamical system, the region
remains invariant---that is, no part of it ``escapes'' under time
evolution. Formally, this condition is expressed as:
\begin{equation}
\bar{X}_{0}^{T}(U)=X_{0}.
\end{equation}

In this formulation, a closed set corresponds to a dynamically stable
region---one that is preserved under the time evolution dictated
by the system $U$. Thus, the classical topological definition of
a closed set gains a physical interpretation: it is a region from
which the dynamical flow cannot exit over the time interval $[0,T]$.

As a simple example, consider the set
\begin{equation}
X=\{1,2,3,4\},
\end{equation}

\begin{equation}
X_{0}=\{2,4\},
\end{equation}
and define the dynamics as
\begin{equation}
U(x,1)=x+2,
\end{equation}
with periodic boundary conditions. Then, after one time step, we have:
\begin{equation}
\overline{X}_{0}^{T}=\{2,4\}=X_{0}
\end{equation}
which satisfies the condition for $X_{0}$ to be a closed set under
this dynamical system.

\subsection{Dynamical Interpretation of Interiors and Open Sets}

An initial region $X_{0}\subset X$ may either remain invariant under
time evolution---that is, satisfy the closedness condition $\bar{X}_{0}^{T}(U)=X_{0}$
---or expand over time, so that $X_{0}\subset\bar{X}_{0}^{T}(U)$.
Similarly, the complement of the initial region, denoted $X_{0}^{c}$,
may also either remain invariant $\overline{X_{0}^{c}}^{T}=X_{0}^{c}$
or expand: $X_{0}^{c}\subset\overline{X_{0}^{c}}^{T}(U)$.

When the complement region $X_{0}^{c}$ remains unchanged under the
evolution, that is,
\begin{equation}
\overline{X_{0}^{c}}^{T}=X_{0}^{c},
\end{equation}
we define the original set $X_{0}$ to be an open set with respect
to the dynamics.

The interior of a set $X_{0}$, denoted $X_{0}^{\circ}$, is then
defined by taking the complement of the closure of its complement:
\begin{equation}
X_{0}^{\circ}\equiv\left(\overline{X_{0}^{c}}^{T}\right)^{c}.
\end{equation}

This definition aligns with classical topology but is now grounded
in a dynamical interpretation: a set is open if the flow does not
``intrude'' into its complement over the time interval considered.

As a concrete example, let us revisit the case from the previous section
involving a closed set. Suppose

\begin{equation}
X=\{1,2,3,4\},
\end{equation}

\begin{equation}
X_{0}=\{2,4\}.
\end{equation}

Then its complement is

\begin{equation}
X_{0}^{c}=\{1,3\},
\end{equation}

and the closure of this complement under the dynamics is

\begin{equation}
\overline{X^{c}}_{0}^{T}=\{1,3\}
\end{equation}

which remains unchanged. Therefore, $X_{0}$ satisfies the condition
for being an open set.

The interior of $X_{0}$\LyXZeroWidthSpace , according to the definition,
is: 
\begin{equation}
X_{0}^{o}=\left(\overline{X_{0}^{c}}^{T}\right)^{c}=\{2,4\}.
\end{equation}

Let us now consider another example. Let

\begin{equation}
X=\{1,2,3,4,5\},
\end{equation}

\begin{equation}
X_{0}=\{1,2,3\},
\end{equation}
and define the time evolution as:

\begin{equation}
U(x,1)=x+1,
\end{equation}
with periodic boundary conditions $U(5,1)=1$.

Then the complement is
\begin{equation}
X_{0}^{c}=\{4,5\},
\end{equation}
and its closure under one time step is:

\begin{equation}
\overline{X_{0}^{c}}^{T}=\{1,4,5\}.
\end{equation}

Since this differs from $X_{0}^{c}$, $X_{0}$ is not an open set
under this dynamical system.

Its interior is given by:
\begin{equation}
X_{0}^{o}=\left(\overline{X_{0}^{c}}^{T}\right)^{c}=\{2,3\}.
\end{equation}

\color{black}
\subsection{Latent Spaces and Stabilization}

The preceding discussion suggests that a topological structure need not be
presupposed from the beginning. Rather, it may be understood as a stabilized
form of a more primitive dynamical structure. In order to make this point
explicit, we introduce the notion of a \emph{latent space}.

Let \(X\) be a set equipped with a dynamical evolution \(U\), and let \(T\)
be a time horizon. The reachability operation
\[
\operatorname{cl}^{T}_{U}(A)
\]
assigns to each subset \(A \subset X\) the region dynamically reachable from
\(A\) within the time interval \(T\). This operation behaves like a closure
operation in the sense that it enlarges a given region according to the
dynamics. However, it need not be idempotent. In general, one may have
\[
\operatorname{cl}^{T}_{U} \circ \operatorname{cl}^{T}_{U}
\neq
\operatorname{cl}^{T}_{U}.
\]
This means that, even after taking the reachability domain once, applying the
same reachability operation again may further enlarge the region. In such a
case, the spatial structure has not yet stabilized. We call this kind of
not-yet-stabilized structure a \emph{latent space}. Mathematically, it is
naturally described as a pre-topological structure: a closure-like structure in
which idempotency is not required.

By contrast, when
\[
\operatorname{cl}^{T}_{U} \circ \operatorname{cl}^{T}_{U}
=
\operatorname{cl}^{T}_{U},
\]
the reachability operation becomes stable under repeated application. Once a
region has been closed under the dynamics, no further enlargement occurs by
applying the same operation again. In this case, the resulting structure may be
regarded as a stabilized topological structure.

Thus, the distinction between pre-topology and topology can be interpreted
dynamically. A pre-topological structure expresses a stage in which reachability
domains are still changing or expanding, whereas a topological structure
expresses a stabilized configuration of reachability. \color{black}Here ``non-stabilized'' and ``stabilized'' describe whether reapplying the reachability operator can further enlarge a region. They do not identify pre-topology with physical nonequilibrium or topology with thermodynamic equilibrium.\color{black}~Rather, the present framework uses idempotency as a mathematical expression of stabilization.

This interpretation also clarifies the transition to the next section. Dynamics
first produces reachability domains and hence a latent spatial structure.
Observation then selects which of these dynamically generated regions are to be
distinguished. The measurable structure discussed below is therefore not imposed
on a pre-given space, but arises from the interaction between a dynamically
generated latent space and an observational context.
\color{black}

\color{black}
\subsection{Dynamical Interpretation of Topological Spaces}

If
\begin{equation}
\operatorname{cl}_U^T(\operatorname{cl}_U^T(A)) =\operatorname{cl}_U^T(A)
\qquad(\forall A\subseteq X)
\end{equation}
holds, the reachability operator is a topological closure operator. In the resulting topology, a closed set $X_0$ satisfies
\begin{equation}
\operatorname{cl}_U^T(X_0)=X_0
\end{equation}
and open sets are the complements of closed sets.

That is, a subset is considered closed if it remains invariant under
the system's time evolution. Taking complements leads to a corresponding
definition of open sets. In both cases, the defining property is that
the evolution does not ``leak out'' of the subset or into its complement,
respectively.

\color{black}
Here stabilization means that reapplying the reachability operator does not enlarge a reachability domain already obtained. It does not require individual points to cease evolving. In this sense, a topological structure is obtained when the reachability operation becomes idempotent.
\color{black}


\color{black}
We could naturally proceed to redefine notions
such as continuity and homeomorphism between pre-topological spaces
in dynamical terms. However, since the focus of this paper is on the
constructive origin of space itself, we omit such technical developments
here and refer interested readers to \cite{key-2} for further elaboration.

\color{black}

\section{Measurable Structures and the Act of Distinction in Observation}

In this section, we explore how measurable structures can be understood
as arising from projective structures associated with observation.
That is, a measurable structure is not something inherently present
in space, but rather reflects the observer's decision about what distinctions
are to be made within that space.

From this viewpoint, observation is the act of instituting distinctions:
it is the selection of criteria by which one portion of a space is
separated from another. The measurable structure, then, encodes the
observer's choice of which features or properties are to be considered
relevant and discernible.

Thus, we propose that measurement is not merely the extraction of
information from a pre-existing structure, but a constructive act
that determines the structure itself---by selecting what distinctions
to recognize, the observer effectively determines the measurable geometry
of the space.

\color{black}
\subsection{From Reachability Domains to Measurable Structures}

In the previous section, we introduced reachability domains and interpreted
the associated dynamical closure operation as giving rise to pre-topological,
and in stable cases topological, structures. We now explain how such
dynamically generated structures are connected to measurable structures.

Let \(X\) be a set equipped with a dynamical evolution \(U\) and a time
horizon \(T\). For each subset \(A \subset X\), we denote by
\(\overline{A}^{\,T}(U)\) the reachability domain generated from \(A\) under
\(U\) up to time \(T\). Equivalently, using the notation introduced in the
previous section, this can be written as \(cl_U^T(A)\). We define the family
of dynamically generated reachability domains by
\[
\mathcal{R}_{U,T}
=
\{\, \overline{A}^{\,T}(U) \mid A \subset X \,\}
=
\{\, cl_U^T(A) \mid A \subset X \,\}.
\]
This family represents the collection of regions made accessible, or
dynamically distinguished, by the evolution \(U\) within the time scale \(T\).

However, an observer does not necessarily distinguish all elements of
\(\mathcal{R}_{U,T}\). What can actually be distinguished depends on the
physical measurement context, including the apparatus, resolution, interaction
with the system, choice of observable, and formation of records. We denote
such a measurement context by \(M\). The context \(M\) selects a subfamily
\[
\mathcal{O}_M \subseteq \mathcal{R}_{U,T},
\]
whose elements are the regions that are physically distinguishable within that
context.\color{black}The context $M$ is specified as a physically constrained input, involving the apparatus, resolution, interaction, and formation of records; it determines which distinctions in $\mathcal O_M$ can be realized. Having specified this relation, we construct $\mathcal B_M=\sigma(\mathcal O_M)$ without requiring the selection and stabilization of $M$ to be derived from $U$ alone. The context $M$ may include a time-dependent measurement axis or procedure.\color{black}

The measurable structure associated with \(M\) is then defined as the
\(\sigma\)-algebra generated by this selected family:
\[
\mathcal{B}_M = \sigma(\mathcal{O}_M).
\]
Thus, the family \(\mathcal{R}_{U,T}\) should not be regarded as a
\(\sigma\)-algebra by itself. Rather, it provides dynamically generated
candidates for distinguishable regions, while the measurement context selects
those distinctions that are actually available to the observer. The measurable
structure is obtained only after this selection and subsequent closure under
the operations required of a \(\sigma\)-algebra.

In this way, the connection between dynamical evolution and measurable
structure can be summarized as
\color{black}
\[
(X,U,T)
\longrightarrow
\mathcal{R}_{U,T},\qquad
(\mathcal{R}_{U,T}, M)
\longrightarrow
\mathcal{O}_M
\longrightarrow
\mathcal{B}_M .
\]
\color{black}
Dynamics generates possible distinctions; measurement realizes actual
distinctions; and the realized distinctions generate the measurable structure.
\color{black}

\color{black}
\subsection{Measurable Spaces Generated by Distinctions}
 We now describe the general mechanism by which the selected family \(\mathcal{O}_M\) generates a measurable structure. The point of the following discussion is not to introduce an arbitrary partition of \(X\), but to clarify how physically available distinctions, selected by the measurement context \(M\), give rise to a \(\sigma\)-algebra. Let \(X\) be a given set. Suppose that, within a measurement context \(M\), a property or observational criterion is assigned to each element \(x \in X\) by means of a binary function \[ p(x) = \begin{cases} 1 & \text{if } x \text{ satisfies the property,} \\ 0 & \text{otherwise.} \end{cases} \] This function represents one distinction that is available within the context \(M\). Equivalently, it determines a pair of distinguishable regions \[ A = p^{-1}(1), \qquad A^c = p^{-1}(0). \] In the terminology of the previous subsection, such regions are to be understood as elements, or Boolean combinations of elements, selected from \(\mathcal{O}_M\). Thus, the construction below describes how the selected distinctions encoded in \(\mathcal{O}_M\) generate the measurable structure \(\mathcal{B}_M = \sigma(\mathcal{O}_M)\).
\color{black}

To refine the distinctions within the space, we may consider multiple
properties, denoted $p_{1},p_{2},\dots,p_{n}$. Each property $p_{i}$
gives rise to a corresponding measurable set:
\begin{equation}
A_{i}=p_{i}^{-1}(1).
\end{equation}
Each $A_{i}$ is the preimage of $1$ under the projection $p_{i}$,
representing the set of points in $X$ for which the $i$-th property
holds.

Each set $A_{i}=p_{i}^{-1}(1)$, corresponding to a particular property
$p_{i}$, can stand in various logical relationships to other sets
$A_{j}=p_{j}^{-1}(1)$ defined by other properties $p_{j}$. If we
consider, for example, just two properties $p_{1}$ and $p_{2}$,
then we can form combinations of the corresponding sets to reflect
various observational queries about the system.

These combinations form a collection $\mathcal{B}$ of subsets of
$X$, given by:
\begin{equation}
\mathcal{B}=\{\slashed{O},A_{1},A_{2},A_{1}^{c},A_{2}^{c},A_{1}\cap A_{2},A_{1}\cup A_{2},A_{1}^{c}\cap A_{2},A_{1}\cap A_{2}^{c},\dots,X\},
\end{equation}
where $\slashed{O}$ is the empty set. These subsets encode all the
possible distinctions that can be made by combining the two properties.

This collection $\mathcal{B}$ is called a $\sigma$-algebra(sigma-algebra),
meaning that it is closed under the operations of complement, countable
union, and countable intersection.

A pair $(X,\mathcal{B}$), where $\mathcal{B}$ is $\sigma$-algebra
of subsets of $X$, is called a measurable space\footnote{The point here is not the textbook definition of a $\sigma$-algebra per se, but the interpretation of this structure as arising from distinctions instituted through acts of observation.}. In our context,
this $\sigma$-algebra reflects the observer's selection of which
distinctions (i.e., which properties) are relevant and observable. 

As a concrete example, consider the finite set:
\begin{equation}
X=\{1,2,3,4,5\}.
\end{equation}
Let us define two properties to distinguish elements of $X$:
\begin{itemize}
\item $p_{e}(x)=1$ if $x$ is even, and $0$ otherwise;
\item $p_{o}(x)=1$ if $x$ is prime number, and 0 otherwise.
\end{itemize}
Then, the corresponding subsets defined by these properties are:

\begin{equation}
A_{e}=p_{e}^{-1}(1)=\{2,4\},\ A_{e}^{c}=p_{e}^{-1}(0)=\{1,3,5\},
\end{equation}

\begin{equation}
A_{p}=p_{p}^{-1}(1)=\{2,3,5\},\ A_{p}^{c}=p_{p}^{-1}(0)=\{1,4\}.
\end{equation}
From these, we can generate a $\sigma$-algebra $\mathcal{B}$ by
forming all Boolean combinations of $A_{e}$ and $A_{p}$. The resulting
collection includes: 
\color{black}
Specifically, the $\sigma$-algebra generated by these distinctions consists of the following sixteen sets:
\[
\begin{aligned}
\mathcal B=\{&
\varnothing,\,
A_e,\,
A_e^c,\,
A_p,\,
A_p^c,\,
A_e\cup A_p,\,
A_e\cap A_p,\\
&
A_e^c\cup A_p,\,
A_e\cup A_p^c,\,
A_e^c\cup A_p^c,\,
A_e^c\cap A_p^c,\,
A_e^c\cap A_p,\,
A_e\cap A_p^c,\\
&
(A_e\cap A_p)\cup(A_e^c\cap A_p^c),\\
&
(A_e\cap A_p^c)\cup(A_e^c\cap A_p),\,
X
\}\\[2mm]
=\{&
\varnothing,\,
\{2,4\},\,
\{1,3,5\},\,
\{2,3,5\},\,
\{1,4\},\\
&
\{2,3,4,5\},\,
\{2\},\,
\{1,2,3,5\},\,
\{1,2,4\},\\
&
\{1,3,4,5\},\,
\{1\},\,
\{3,5\},\,
\{4\},\\
&
\{1,2\},\,
\{3,4,5\},\,
\{1,2,3,4,5\}
\}.
\end{aligned}
\]
\color{black}

Each of these subsets corresponds to a particular distinction made
by combining the properties of evenness and primality. For instance,
the set of elements that are even but not prime is given by:
\begin{equation}
A_{e}\cap A_{p}^{c}=\{4\}.
\end{equation}
This illustrates how observational distinctions---based on multiple
binary properties---can generate a $\sigma$-algebra, and hence a
measurable structure, on a given space.

\color{black}
\subsection{Topological and Pre-Topological Structures as Sources of Distinction}

The preceding discussion can now be related to the dynamical structures
introduced in Section 2.
\color{black}
When the reachability operator is idempotent, let $\tau$ denote the topology that it defines. If all open sets are adopted as observational distinctions, then
\begin{equation}
\mathcal B_M=\sigma(\tau)
\end{equation}
is the Borel $\sigma$-algebra of this topology. If the measurement context further restricts the available distinctions, the corresponding measurable structure reflects that restriction.
\color{black}

However, the present framework is not restricted to this stable case. In a
pre-topological situation, the reachability domain \(cl_U^T(A)\) need not
coincide with the initial region \(A\). Even in this case, the family of
reachability domains
\[
\mathcal{R}_{U,T}
=
\{\,cl_U^T(A)\mid A\subset X\,\}
\]
can provide candidates for distinguishable regions. A measurement context
\(M\) then selects a subfamily
\[
\mathcal{O}_M\subseteq \mathcal{R}_{U,T},
\]
and the associated measurable structure is given by
\[
\mathcal{B}_M=\sigma(\mathcal{O}_M).
\]

Thus, topological spaces appear as stable cases in which dynamically generated
regions no longer expand under the chosen time scale, while pre-topological
structures describe more general situations in which reachability domains remain
dynamically generated and context-dependent. In both cases, measurable
structures arise not from space alone, but from the combination of dynamical
reachability and measurement-based distinction.
\color{black}

\subsection{Introducing Measures: The Foundation of the Concept of State}

Thus far, we have treated space as something that emerges dynamically:
a structure shaped by the time evolution of a system, represented
through reachability domains. However, all of this structure has been
kinematic---it tells us what regions are dynamically possible, but
says nothing about how likely each possibility is to occur.

In this view, a topological space encodes which trajectories are possible
under the system's dynamics. But to describe the actual behavior of
a system---or more precisely, its state---we must specify how these
possibilities are weighted. This is where the notion of measure enters:
it assigns a numerical value to each region, quantifying how much
the system is expected to occupy or pass through that region.
\begin{equation}
\mu:\mathcal{B}\to[0,1],\ \mu(X_{0})\in\mathbb{R}_{\geq0}.
\end{equation}
When interpreted probabilistically, the entire space satisfies

\begin{equation}
\mu(X)=1,
\end{equation}
and the measure defines a state of the system. The key insight is
that the state is not an intrinsic object, but arises from assigning
possibility weights to dynamically accessible regions. In this sense,
the measure is not something added on top of space---it is the quantitative
expression of the system's relationship to its space.

\color{black}
Represent an observable on the measurable space $(X,\mathcal B)$ by a measurable function
\begin{equation}
  \hat{x}:X\longrightarrow\mathbb R
\end{equation}
The value of the observable at $\xi\in X$ is denoted by $x=\hat{x}(\xi)$.

Let the state be given by a probability measure $\mu$ on $X$. If $f(\hat{x})$ is integrable, its expectation is
\begin{equation}
  \omega_\mu\bigl(f(\hat{x})\bigr)
  =
  \int_X f\bigl(\hat{x}(\xi)\bigr)\,d\mu(\xi)
  =
  \int_{\mathbb R} f(x)\,d\mu_{\hat{x}}(x)
\end{equation}
Here $\mu_{\hat{x}}$ is the probability distribution of the measurement outcome, defined by
\begin{equation}
  \mu_{\hat{x}}(\Delta)
  =
  \mu\bigl(\hat{x}^{-1}(\Delta)\bigr),
  \qquad
  \Delta\in\operatorname{Borel}(\mathbb R)
\end{equation}
If $\mu_{\hat{x}}$ admits a probability density $\rho(x)$, the expectation can also be written as
\begin{equation}
  \omega_\mu\bigl(f(\hat{x})\bigr)
  =
  \int_{\mathbb R}f(x)\rho(x)\,dx
\end{equation}
.
\color{black}

The triple $(X,\mathcal{B},\mu)$ is thus called a measure space.
Within this framework, we can now clearly state:

A state is a map that assigns a distribution of possibility (i.e.,
a measure) to a structure of observation (i.e., the $\sigma$-algebra
$\mathcal{B}$).

\color{black}
The measurement context determines which observables and events can be distinguished. The state assigns probabilities to those events. Since different states may be defined on the same measurable structure, the choice of measurement context alone does not uniquely determine the probability distribution.
\color{black}

\color{black}
\subsection{Reachability Dynamics and Observable Dynamics}

Before turning to the noncommutative case, it is useful to clarify that the
term ``dynamics'' is used in two related but distinct senses in the present
framework. This distinction is important because the dynamical construction of
reachability domains and the time evolution of observables do not play the same
role.

First, there is the dynamics of reachability. This is the dynamics considered
in Section 2. It is given by a family of maps
\[
U_t : X \to X,
\]
which evolves points, states, or regions of the underlying set \(X\). For an
initial region \(A \subset X\), this dynamics generates the reachability domain
\[
cl_U^T(A)
=
\{\, U_t(x) \mid x \in A,\ 0 \leq t \leq T \,\}.
\]
Thus, the role of \(U_t\) is to generate dynamically accessible regions and,
through the associated closure operation, to induce a pre-topological structure.
When the closure operation becomes idempotent, this structure may be regarded
as a stabilized topological structure. In the categorical formulation developed
elsewhere, this layer corresponds to a time-evolution functor
\[
F(T):\mathcal{C}\to\mathcal{C},
\]
whose action plays the role of a closure operation generating what we call a
latent space or pre-topological space.

\color{black}
Next, consider a function $f:X\to\mathbb C$ on the state set $X$. The time evolution $U_t:X\to X$ acts on $f$ by
\begin{equation}
\alpha_t(f)=f\circ U_t
\end{equation}
Thus the same evolution is described by $U_t$ on states and by $\alpha_t$ on functions.

No topology on $X$ needs to be presupposed to define this action on functions. In this paper, a topology arises when the reachability operator $\operatorname{cl}_U^T$ is idempotent and therefore defines a topological closure operator.
\color{black}

In quantum theory, however, a point-set space \(X\) is not assumed as a
primitive object. The basic structure is instead a generally noncommutative
algebra of observables \(\mathcal{A}\), together with a time evolution
\[
\alpha_t:\mathcal{A}\to\mathcal{A}.
\]
In the Heisenberg picture, for example, this evolution is typically written as
\[
\alpha_t(a)=e^{itH}ae^{-itH},
\]
where \(H\) is the Hamiltonian and we set \(\hbar=1\). In this setting, the
observable dynamics \(\alpha_t\) is primary. Spatial or measurable structures
are not given in advance, but are reconstructed locally through the choice of a
measurement context.

The distinction between the two layers of dynamics may be summarized as
follows:
\[
\begin{array}{c|c|c}
\text{Type of dynamics} & \text{Notation} & \text{Role} \\
\hline
\text{Reachability dynamics} & U_t:X\to X
& \text{generates reachability domains and pre-topological structures} \\
\text{Observable dynamics} & \alpha_t:\mathcal{A}\to\mathcal{A}
& \text{describes the time evolution of observables}
\end{array}
\]
In the classical case, these two layers are related by duality:
\[
U_t:X\to X
\quad\Longrightarrow\quad
\alpha_t=U_t^{*},
\qquad
U_t^{*}(f)=f\circ U_t.
\]
In the quantum case, by contrast, the observable dynamics
\(\alpha_t:\mathcal{A}\to\mathcal{A}\) is taken as fundamental. A measurement
context \(M\) then selects a commutative subalgebra
\[
\mathcal{C}_M \subset \mathcal{A},
\]
and from this selected context a local measurable structure is constructed.
Thus, the transition from classical to quantum theory can be understood, within
the present framework, as a transition from a globally available space \(X\) to
context-dependent measurable structures generated from commutative parts of a
noncommutative observable algebra.
\color{black}

\subsection{Noncommutative Space and Quantum Theory}

The construction of a measure space, as described above, naturally
extends to cases where the observables are noncommutative---that
is, when they cannot be simultaneously measured or diagonalized. The
technical details of this extension are deferred to Appendix A, where
we describe how a Hilbert space arises via the GNS (Gelfand--Naimark--Segal)
construction once a state and a dynamical algebra are given.

From the perspective of quantum systems, the essential difference
between classical and quantum theories lies in the commutativity of
observables. In classical theory, all observables commute, and a single
global measure space suffices to describe the system. In contrast,
in quantum theory, observables may not commute, and only subsets of
commuting observables can be simultaneously assigned measurable structure.

When an observer selects a particular observable $\hat{x}$, only
those observables that commute with $\hat{x}$ can meaningfully define
a simultaneous measurement context. The corresponding measure space
$(X(\hat{x}),\mathcal{B}(\hat{x}),\mu(\hat{x}))$ emerges as the projection
of the system onto this selected observable context. In other words,
the ``space'' that becomes visible to the observer is the one constructed
by the act of choosing $\hat{x}$.

Thus, even in the quantum case, spatial structure remains fundamentally
tied to the observer's act of selection. What appears as ``space''
is not a fixed background but a context-dependent construct---one
shaped by the algebraic properties of observables and the epistemic
stance of the observer.

\color{black}
In this setting, the role of the GNS construction should be understood
as follows. It is not introduced merely as an abstract representation
theorem, but as the mathematical mechanism by which a state on an algebra
of observables gives rise to a representation space in which measurement
contexts can be analyzed. More precisely, given a pair
\((\mathcal{A},\omega)\), where \(\mathcal{A}\) is an algebra of
observables and \(\omega\) is a state on it, the GNS construction yields
a triple
\[
(\mathcal{H}_{\omega},\pi_{\omega},\Omega_{\omega}).
\]
Within this representation, the choice of a measurement context \(M\)
amounts to selecting a commutative subalgebra
\[
\mathcal{C}_{M} \subset \pi_{\omega}(\mathcal{A}).
\]
The spectrum of this commutative algebra then provides the underlying
set of possible outcomes,
\[
X_{M} := \operatorname{Spec}(\mathcal{C}_{M}),
\]
and the state \(\omega\) induces, by restriction to \(\mathcal{C}_{M}\),
a probability measure \(\mu_{M}\) on the corresponding measurable
structure \(\mathcal{B}_{M}\).
\color{black}Here $X_M$ is the outcome space associated with the selected commutative context; this construction alone does not identify it with physical space or spacetime.\color{black}

Thus, the relevant chain of construction
can be summarized as
\[
(\mathcal{A},\omega)
\longrightarrow
(\mathcal{H}_{\omega},\pi_{\omega},\Omega_{\omega})
\longrightarrow
\mathcal{C}_{M}
\longrightarrow
(X_{M},\mathcal{B}_{M},\mu_{M}).
\]
In this sense, the GNS construction serves as the bridge from a
state-dependent noncommutative algebraic description to the
context-dependent measurable spaces that appear through observational
selection.
\color{black}

The constructive nature of spatial structure---and its noncommutative
generalization---are elaborated in detail in Appendix A.

\subsection{Constructive Dynamics and Measurement in Spin Precession}

As a concrete example, we consider the precession of a spin in an
external magnetic field. We demonstrate how the dynamical evolution
of the spin gives rise to a reachability domain, and how the observer's
choice of observable affects not only the associated measurable structure
but also the reachability domain itself\footnote{While the analysis of spin precession is standard, our discussion differs in that it interprets the resulting trajectories as reachability domains whose structure depends on the choice of observable, thereby providing a concrete example of space emerging from dynamics and measurement.}.

When spin is present, its interaction with an external magnetic field
is described by the Hamiltonian

\begin{equation}
\hat{H}_{I}=-\hat{\mu}_{i}B^{i}
\end{equation}
\foreignlanguage{english}{where $\mu_{i}$ denotes the magnetic moment
of the particle, and $B^{i}$ represents the external magnetic field.}

For an electron, the spin-induced magnetic moment is given by

\begin{equation}
\hat{\mu}_{i}=-g\mu_{B}\frac{\hat{\psi}_{i}}{\hbar}
\end{equation}
where $g\approx2$ is the $g$-factor for a free electron, and $\mu_{B}=\frac{e\hbar}{2m_{e}}$
is the Bohr magneton.

The operator $\hat{\psi}_{i}$ is defined by

\begin{equation}
\hat{\psi}_{i}=\frac{\hbar\sigma_{i}}{2}
\end{equation}
where $\sigma_{i}$ are the Pauli matrices, satisfying the commutation
and anticommutation relations

\begin{equation}
[\sigma_{i},\ \sigma_{j}]=2i\epsilon_{ijk}\sigma_{k},\ \{\sigma_{i},\sigma_{j}\}=2\delta_{ij}.
\end{equation}
From this, the corresponding commutation and anticommutation relations
for $\hat{\psi}_{i}$ are

\[
[\hat{\psi}_{i},\ \hat{\psi}_{j}]=i\hbar\epsilon_{ijk}\hat{\psi}_{k},\ \{\hat{\psi}_{i},\hat{\psi}_{j}\}=\frac{\hbar^{2}}{2}\delta_{ij}.
\]

In the literature, the spin operator $\hat{\psi}_{i}$ is often denoted
as $\hat{S}_{i}$.

Now, we consider a situation in which the particle is either extremely
massive or subject to a strong external magnetic field, such that
the free Hamiltonian $\hat{H}_{0}$ can be neglected. That is, we
approximate

\begin{equation}
\color{black}\hat{H}=\hat{H}_{0}+\hat{H}_{I}\approx-\hat{\mu}_{i}B^{i}=g\frac{e}{2m_{e}}\hat{\psi}_{i}B^{i}=g\frac{e\hbar B^{i}}{4m_{e}}\sigma_{i}.\color{black}
\end{equation}
We now consider the time evolution of the spin observable $\hat{\psi}_{i}$.
In the Heisenberg picture, the time-dependent operator is given by

\begin{equation}
\hat{\psi}_{i}(t)=\hat{U}^{\dagger}(t)\hat{\psi}_{i}\hat{U}(t),
\end{equation}
where the unitary evolution operator is

\begin{equation}
\color{black}\hat{U}(t)=e^{-i\frac{\hat{H}t}{\hbar}}=\exp\left(-ig\frac{et}{2m_{e}\hbar}\hat{\psi}_{i}B^{i}\right)=\exp\left(-i\frac{geB^{i}}{4m_{e}}t\sigma_{i}\right).\color{black}
\end{equation}
\foreignlanguage{english}{We define
\begin{equation}
\tilde{B}^{i}\equiv\frac{geB_{i}}{4m_{e}},\ \tilde{B}\equiv|\tilde{\bm{B}}|,\ n^{i}\equiv\frac{\tilde{B}^{i}}{\tilde{B}},
\end{equation}
so that the unitary operator becomes
\begin{equation}
\color{black}\hat{U}(t)=\exp\left(-i\tilde{B}^{i}t\sigma_{i}\right)=I\cos\left(\tilde{B}t\right)-in^{i}\sigma_{i}\sin\left(\tilde{B}t\right).\color{black}
\end{equation}
}Since the unitary action of Pauli matrices corresponds to a rotation
by twice the angle in vector space, we have

\begin{equation}
\hat{\psi}_{i}(t)=\hat{U}^{\dagger}(t)\hat{\psi}_{i}\hat{U}(t)=R^{ij}(2\tilde{B}t)\hat{\psi}_{j}
\end{equation}
\foreignlanguage{english}{where $R^{ij}(\theta)$ is the $\mathrm{SO}(3)$
rotation matrix by angle $\theta$ around axis $\vec{n}$.}

The infinitesimal time evolution is then governed by the Heisenberg
equation of motion:

\begin{equation}
\color{black}i\hbar\frac{d\hat{\psi}_{i}(t)}{dt}=[\hat{\psi}_{i}(t),\hat{H}]=g\frac{e}{2m_{e}}[\hat{\psi}_{i}(t),B^{j}\hat{\psi}_{j}(t)].\color{black}
\end{equation}
Using the commutation relations, this becomes

\begin{equation}
\color{black}[\hat{\psi}_{i}(t),B^{j}\hat{\psi}_{j}(t)]=i\hbar\epsilon_{ijk}B^{j}\hat{\psi}_{k}(t),\color{black}
\end{equation}
and we obtain

\begin{equation}
\color{black}\frac{d\hat{\psi}_{i}(t)}{dt}=g\frac{e}{2m_{e}}\epsilon_{ijk}B^{j}\hat{\psi}_{k}(t).\color{black}
\end{equation}
This is the precession equation for the spin operator in a magnetic
field.

In the Schr$\ddot{\mathrm{o}}$dinger picture, we introduce the spin
ladder operators

\begin{equation}
\hat{\psi}_{\pm}=\frac{1}{\sqrt{2}}(\hat{\psi}_{1}\pm i\hat{\psi}_{2})
\end{equation}
which satisfy $\hat{\psi}_{-}|0\rangle=0$ where $|0\rangle$ denotes
the spin-down state.

The time-evolved state vector is given by

\begin{equation}
\color{black}|\Psi(t)\rangle=\hat{U}(t)|0\rangle=\cos(\tilde{B}t)|0\rangle-i\sin(\tilde{B}t)\left(\frac{2}{\hbar}n^{i}\hat{\psi}_{i}\right)|0\rangle.\color{black}
\end{equation}
This can be further decomposed using the basis $|0\rangle$, $|1\rangle$
as

\begin{equation}
\color{black}\begin{aligned}|\Psi(t)\rangle= & \left(\cos(\tilde{B}t)-i\sin(\tilde{B}t)\frac{2}{\hbar}n^{i}\langle0|\hat{\psi}_{i}|0\rangle\right)|0\rangle\\
 & -i\sin(\tilde{B}t)\frac{2}{\hbar}n^{i}\langle1|\hat{\psi}_{i}|0\rangle|1\rangle.
\end{aligned}\color{black}
\end{equation}
Since the spin operators are related to the Pauli matrices via

\begin{equation}
\hat{\psi}_{i}=\frac{\hbar\sigma_{i}}{2},
\end{equation}
we adopt the standard representation of the Pauli matrices:

\begin{equation}
\sigma_{1}=\left(\begin{array}{cc}
0 & 1\\
1 & 0
\end{array}\right),\ \sigma_{2}=\left(\begin{array}{cc}
0 & -i\\
i & 0
\end{array}\right),\ \sigma_{3}=\left(\begin{array}{cc}
1 & 0\\
0 & -1
\end{array}\right).
\end{equation}
Under this representation, the time-evolved spin state can be written
as a two-component spinor on the Bloch sphere.
\color{black}The column vector below uses the component order $(|1\rangle,|0\rangle)$, with $|0\rangle=(0,1)^{\mathsf T}$.\color{black}
\begin{equation}
\color{black}|\Psi(t)\rangle=\left(\cos\left(\tilde{B}t\right)+in^{3}\sin(\tilde{B}t)\right)|0\rangle-i\sin\left(\tilde{B}t\right)(n^{1}-in^{2})|1\rangle=\left(\begin{array}{c}
-i(n^{1}-in^{2})\sin\left(\tilde{B}t\right)\\
\cos\left(\tilde{B}t\right)+in^{3}\sin(\tilde{B}t)
\end{array}\right).\color{black}
\end{equation}
One can verify that this state remains normalized:

\begin{equation}
\langle\Psi(t)|\Psi(t)\rangle=\cos^{2}\left(\tilde{B}t\right)+\sin^{2}\left(\tilde{B}t\right)=1.
\end{equation}
Thus, the trajectory of the spin state under time evolution describes
a circular motion on the Bloch sphere, centered around the axis defined
by the unit vector $\vec{n}$.

Since the trajectory traced by the state vector is a closed circle
on the Bloch sphere, the reachability domain for times $t>\frac{2\pi}{\tilde{B}}$
becomes the circle $S^{1}$ around the axis defined by the vector
$\vec{n}$.

Now consider the initial state $|0\rangle$, which satisfies
\begin{equation}
\hat{\psi}_{3}|0\rangle=-\frac{\hbar}{2}|0\rangle.
\end{equation}
Using the time evolution operator, we find

\begin{equation}
\hat{U}(t)\hat{\psi}_{3}\hat{U}(t)^{-1}\hat{U}(t)|0\rangle=-\frac{\hbar}{2}\hat{U}(t)|0\rangle,
\end{equation}
\color{black}
Here the co-rotating observable is defined by
\begin{equation}
\hat{\psi}_{3,\mathrm{co}}(t)=\hat{U}(t)\hat{\psi}_{3}\hat{U}^{\dagger}(t),
\end{equation}
which is distinct from the Heisenberg-picture operator
$\hat{\psi}_{3}(t)=\hat{U}^{\dagger}(t)\hat{\psi}_{3}\hat{U}(t)$
defined above. The preceding equation therefore implies
\begin{equation}
\hat{\psi}_{3,\mathrm{co}}(t)|\Psi(t)\rangle=-\frac{\hbar}{2}|\Psi(t)\rangle.
\end{equation}
Thus the measurement outcome along the co-rotating axis remains
$-\hbar/2$ at all times, and the state remains an eigenstate of the
co-rotating observable.
\color{black}

In contrast, if we do not adjust the observable to follow the spin's
precession and simply measure $\hat{\psi}_{3}$, the system evolves
into different superpositions over time, and after a full cycle $t=\frac{2\pi}{\tilde{B}}$,
the state traverses the reachability domain $S^{1}$.

Similarly, if we instead choose $\hat{\psi}_{2}$ as the observable,
the state is projected into an eigenstate of $\hat{\psi}_{2}$ at
$t=0$, and its subsequent time evolution generates a different reachability
domain $S'{}^{1}$, which is another great circle on the Bloch sphere
rotated around the same axis $\vec{n}$.

The orientation---or inclination---of this reachability domain with
respect to the Bloch sphere is determined by the choice of observable,
and thus reflects the structure of measurability imposed by the observer.

\subsubsection{3.7.1 Interpretation: Reachability Domains as Constructed Spaces}

In summary, while the precessional motion of spin on the Bloch sphere
is well known as a unitary trajectory in the conventional picture,
our approach interprets this trajectory not merely as an abstract
path in state space, but as a dynamically generated reachability domain---a
spatial structure that becomes accessible through observational acts.
Unlike static geometric backgrounds, this domain reflects the concrete
interface between system and observer: it is defined by the interplay
between time evolution and the chosen observable. By synchronizing
the observable with the system's dynamics, one constructs an effectively
static spatial frame, whereas fixing the observable yields a dynamically
expanding set of accessible outcomes. Thus, space in this framework
is not presupposed but emerges as a context-dependent construct. This
perspective offers a constructive alternative to classical realism
and aligns naturally with the contextual and noncommutative features
of quantum theory, suggesting a viable foundation for quantum-geometric
approaches to space.

\section{Conclusion and Discussion}

In this paper, we have analyzed and reconstructed the concept of space
through the lens of physical frameworks such as classical and quantum
systems. Our central claim has been that dynamical evolution and observational
acts together serve as fundamental building blocks in the formation
of spatial structure.

\color{black}
This point also clarifies the relation between the present framework and
Rovelli's relational quantum mechanics. Relational quantum mechanics
emphasizes that physical facts are not absolute, but are defined relative to
interactions between physical systems. In this respect, the present framework
is close in spirit to Rovelli's relational view. However, our emphasis is
different. Rather than focusing only on the relational status of physical facts,
we ask how such relationally established facts give rise to measurable and
spatial structures.

In the present framework, the observer is not treated as an abstract
mathematical agent who freely partitions a set. Observation is instead
described through a measurement context
\[
M=(\text{apparatus},\text{resolution},\text{interaction},
\text{chosen observable},\text{record}),
\]
which is physically constrained. This context selects the distinguishable
regions \(\mathcal{O}_M\) from the dynamically generated reachability domains
\(\mathcal{R}_{U,T}\), and the associated measurable structure is given by
\[
\mathcal{B}_M=\sigma(\mathcal{O}_M).
\]
Thus, while relational quantum mechanics stresses the observer-relative
character of physical facts, the present framework focuses on the further
structural question of how observer-relative facts, together with dynamics and
measurement context, generate measurable or spatial structures.
\color{black}

\color{black}
The relation to existing theories of quantum measurement should also be
understood in this limited sense. The present framework does not aim to solve
the measurement problem in the standard sense. Rather, it provides a structural
account of how measurable and spatial structures are generated once a
measurement context is specified. In this respect, decoherence theory may be
seen as describing how certain distinctions become dynamically stable through
interaction with an environment\cite{key-dec}, while Ozawa's theory of quantum measurement
clarifies how measurement error and disturbance constrain which distinctions
can be physically realized\cite{key-ozawa}. Similarly, operational descriptions in terms of
POVMs or instruments may be used to specify the measurement context itself,
including apparatus, interaction, outcomes, and record formation\cite{key-POVM}. These
approaches are therefore complementary to the present framework: they help
specify or constrain the measurement context \(M\), while the present framework
describes how \(M\) selects distinguishable regions \(\mathcal{O}_M\) and gives
rise to the measurable structure \(\mathcal{B}_M=\sigma(\mathcal{O}_M)\).
\color{black}

\color{black}
Finally, the relation between the present framework and Ojima's
micro--macro duality\cite{key-Ojima}\cite{key-Ojima2}, as well as category-theoretic approaches to quantum
foundations such as those developed by Saigo\cite{key-Saigo}\cite{key-Saigo2}, remains an important topic for
future work.
\color{black}

This perspective reveals a deep and intricate connection between the
concept of space in mathematics and that in physics. Rather than treating
space as a fixed, pre-given background, we have shown that it may
emerge from the interplay between system dynamics and the observer's
mode of engagement. This approach invites a rethinking of mathematical
foundations in a way that is more aligned with the operational language
of physics---a reconstruction of mathematics that is, in some sense,
physically meaningful.

However, such a shift also highlights the need for a reconstruction
of mathematics itself, one that moves beyond the static formulations
characteristic of classical mathematics prior to the 20th century.
Developing a framework in which space is not presupposed but dynamically
and observationally constructed remains a major open challenge for
future work.

\section*{Acknowledgments}

I would like to express my gratitude to Associate Professor Shiro
Komata and Dr. Tsukasa Yumibayashi for reading the manuscript and
providing valuable references. I am also indebted to Professor Emeritus
Akio Sugamoto for his insightful comments on discrete and concrete
models. Additionally, I thank the members of the Mathematical Physics
Seminar for their numerous constructive comments on the earlier version
of this work, originally published as \cite{key-2}. These suggestions
have been reflected throughout the present paper. I am grateful to
Kazuya Mitsutani for his helpful comments on the Introduction.

\appendix\numberwithin{equation}{section}

\section{The GNS Construction and the Constructive Understanding of Space}

In this appendix, we consider the formulation of quantum theory as
a dynamical system, focusing on the construction of a corresponding
Hilbert space given a state and a dynamical algebra. This construction
is known as the GNS (Gelfand--Naimark--Segal) representation.

From the perspective of noncommutative geometry, the idea that space
is nothing but algebraic structure has been strongly advocated---see
Connes\cite{key-7} or a foundational treatment. In particular, our
position in this paper---that spatial structure emerges from the
commutative part of observables---can be seen as a noncommutative
generalization of the classical Gelfand--Naimark duality for commutative
$C^{*}$-algebras. An accessible introduction to this viewpoint is
provided in \cite{key-17}.

It is crucial to emphasize here that the definition of space depends
on the observer's choice of observable. The resulting spatial structure
is not absolute, but contingent on the measurement context imposed
by the observer.

Quantum theory, in this sense, can be viewed as a dynamical system
in terms of observables $\hat{x}$. The essential shift from classical
to quantum theory lies in the fact that these observables are no longer
assumed to be mutually commutative. 

To describe the dynamics of a quantum system, we begin by introducing
a $*$-algebra as the algebraic structure of observables. For a rigorous
definition and detailed proofs, see Bratteli and Robinson \cite{key-12}.
In Dirac's terminology, this corresponds to moving from a dynamical
system in the world of c-numbers (classical numbers) to one defined
by q-numbers (quantum operators). Put more intuitively, this is equivalent
to representing observables as complex matrices.In finite-dimensional
systems, such as spin systems, this identification is exact: observables
are literally represented as complex matrices acting on a Hilbert
space.

Let us denote the collection of all observables by $\hat{X}$, with
$\hat{x}\in\hat{X}$. A $*$-algebra is an algebra equipped with an
involutive operation, denoted $\dagger$ (or $*$ in mathematical
notation), which satisfies certain algebraic properties. Specifically,
applying $\dagger$ twice returns the original element:

\begin{equation}
\left(\hat{x}^{\dagger}\right)^{\dagger}=\hat{x}.
\end{equation}

In finite-dimensional systems, this operation corresponds to complex
conjugate transpose---that is, Hermitian adjoint---of matrices:
\begin{equation}
\hat{x}\mapsto\hat{x}^{\dagger}.
\end{equation}
The adjoint operation satisfies the following properties:
\begin{itemize}
\item For linear combinations:
\begin{equation}
\left(a\hat{x}_{1}+b\hat{x}_{2}\right)^{\dagger}=a^{*}\hat{x}_{1}^{\dagger}+b^{*}\hat{x}_{2}^{\dagger},
\end{equation}
where $a^{*}$ is denotes the complex conjugate of $a$.
\item For products:
\begin{equation}
\left(\hat{x}_{1}\hat{x}_{2}\right)^{\dagger}=\hat{x}_{2}^{\dagger}\hat{x}_{1}^{\dagger}.
\end{equation}
\end{itemize}
An observable $\hat{x}$ satisfying $\hat{x}=\hat{x}^{\dagger}$ is
called self-adjoint, and corresponds to a physical quantity that can
be measured.

The time evolution of an observable in quantum mechanics is written
as:
\begin{equation}
\hat{x}(t)=U(\hat{x}_{0},t),\ \mathrm{with\ }U(\hat{x}_{0},0)=\hat{x}_{0}.
\end{equation}

In the case of a Hamiltonian dynamical system, this evolution is typically
given by the Heisenberg picture:
\begin{equation}
\hat{x}(t)=e^{it\hat{H}}\hat{x}_{0}e^{-it\hat{H}},
\end{equation}
where $\hat{H}$ is a self-adjoint Hamiltonian operator generating
the unitary time evolution.

The expectation value of an observable $\hat{x}$ is encoded by a
state functional $\omega$, defined as:

\begin{equation}
\omega(\hat{x})=\langle\hat{x}\rangle.
\end{equation}
We require that $\omega(1)=1$, ensuring normalization. However, since
the observable $\hat{x}$ may not commute with others, this expectation
value cannot always be interpreted directly as a classical measure.
To connect such functionals to Hilbert space structure, we invoke
the GNS (Gelfand--Naimark--Segal) representation.

The GNS construction is a systematic procedure that, given a $*$-algebra
$\mathcal{A}$ and a state $\omega$, produces a Hilbert space representation
in which $\omega$ is realized as an inner product state.

Specifically, consider products of the form $\hat{x}^{\dagger}\hat{y}$.
We interpret this as the inner product between two vectors:

\begin{equation}
\omega(\hat{x}^{\dagger}\hat{y})=\langle\hat{x}|\hat{y}\rangle,
\end{equation}
where $|\hat{y}\rangle$ and $\langle\hat{x}|$ are understood as
elements of a vector space and its dual, respectively.

However, this construction faces a subtle issue: it is possible that

\begin{equation}
\langle\hat{x}|\hat{x}\rangle=\omega(\hat{x}^{\dagger}\hat{x})=0,
\end{equation}
even though $\hat{x}\neq0$ in the algebra. To resolve this, we define
a subspace $\mathcal{N}_{\omega}\subset\mathcal{A}$, consisting of
all null vectors:
\begin{equation}
\mathcal{N}_{\omega}=\{\hat{x}_{\mathrm{null}}\in\mathcal{A}|\omega(\hat{x}^{\dagger}\hat{x})=0\}.
\end{equation}

We then define an equivalence relation:
\begin{equation}
|\hat{x}\rangle\sim|\hat{x}+\hat{x}_{\mathrm{null}}\rangle
\end{equation}
and construct equivalence classes
\begin{equation}
|[\hat{x}]_{\omega}\rangle\in\mathcal{A}/\mathcal{N}_{\omega}.
\end{equation}

A distinguished unit vector is given by the class corresponding to
the identity:

\begin{equation}
|\Omega_{\omega}\rangle=|[1]_{\omega}\rangle\sim|[1+\hat{x}_{\mathrm{null}}]_{\omega}\rangle,
\end{equation}
which satisfies the normalization condition:
\begin{equation}
\langle\Omega_{\omega}|\Omega_{\omega}\rangle=\omega(1)=1.
\end{equation}

Mathematically, the inner product space constructed from the equivalence
classes can be completed to form a Hilbert space, denoted $\mathcal{H}_{\omega}$.

We now define the action of an observable $\hat{x}\in\mathcal{A}$
on this Hilbert space by left multiplication:

\begin{equation}
\pi_{\omega}(\hat{x})|[\hat{y}]\rangle=|[\hat{x}\hat{y}]_{\omega}\rangle.
\end{equation}

This definition turns $\pi_{\omega}$ into a $*$-representation of
algebra $\mathcal{A}$ on the Hilbert space $\mathcal{H}_{\omega}$.
In particular, we recover the equivalence class of $\hat{x}$ itself
as:

\begin{equation}
|[\hat{x}]_{\omega}\rangle=\pi_{\omega}(\hat{x})|[\hat{1}]_{\omega}\rangle.
\end{equation}

Consequently, the expectation value of an observable in the state
$\omega$ can be written as:

\begin{equation}
\omega(\hat{x})=\langle\hat{x}\rangle=\langle\Omega_{\omega}|\pi_{\omega}(\hat{x})|\Omega_{\omega}\rangle=\langle\Omega_{\omega}|[\hat{x}]_{\omega}\rangle,
\end{equation}
where $|\Omega_{\omega}\rangle=|[1]_{\omega}\rangle$ is the distinguished
cyclic vector.

The pair $(\mathcal{H}_{\omega},\pi_{\omega})$ is called the GNS
representation of the state $\omega$. Since the entire Hilbert space
is generated by acting on the cyclic vector $|\Omega_{\omega}\rangle$,
this representation is also known as a cyclic representation. 

Let us now consider a simple example of a $*$-algebra in quantum
mechanics: the quantum harmonic oscillator. This system is defined
using self-adjoint operators $\hat{x}$ and $\hat{p}$ satisfying
the canonical commutation relation:

\begin{equation}
[\hat{x},\hat{p}]=i\hbar.
\end{equation}

\color{black}
For an oscillator of mass $m$ and angular frequency $\omega_{0}$, we introduce
\begin{equation}
\hat{a}=\sqrt{\frac{m\omega_{0}}{2\hbar}}\,\hat{x}
+\frac{i\hat{p}}{\sqrt{2m\hbar\omega_{0}}},\qquad
\hat{a}^{\dagger}=\sqrt{\frac{m\omega_{0}}{2\hbar}}\,\hat{x}
-\frac{i\hat{p}}{\sqrt{2m\hbar\omega_{0}}}.
\end{equation}
\color{black}
which satisfy the fundamental commutation relation:

\begin{equation}
[\hat{a},\hat{a}^{\dagger}]=1
\end{equation}
Any element $\hat{A}\in\hat{X}$ in $*$-algebra of observables can
be written as a (possibly infinite) linear combination:

\begin{equation}
\hat{A}=\sum_{n,m=0}^{\infty}c_{n,m}\hat{a}^{\dagger n}\hat{a}^{m}
\end{equation}

Now consider a state $\omega_{0}$ such that:
\begin{equation}
\omega_{0}(\hat{a}^{\dagger}\hat{a})=0
\end{equation}

This implies:
\begin{equation}
\langle\hat{a}|\hat{a}\rangle=0,
\end{equation}
so the equivalence class of $\hat{a}$ is null: $|\hat{a}\rangle\sim0$.
Then, the cyclic vector becomes:

\begin{equation}
|\Omega_{\omega_{0}}\rangle=|[1]_{\omega_{0}}\rangle\sim|[1+\hat{a}]_{\omega_{0}}\rangle,
\end{equation}
and we have:
\begin{equation}
\pi_{\omega_{0}}(\hat{a})|\Omega_{\omega_{0}}\rangle=|[\hat{a}]_{\omega_{0}}\rangle=0.
\end{equation}
More generally, we can construct a basis of states as:
\begin{equation}
|n\rangle=\frac{1}{\sqrt{n!}}(\pi_{\omega_{0}}(\hat{a}^{\dagger}))^{n}|\Omega_{\omega_{0}}\rangle
\end{equation}

In this way, all physical states are generated by acting repeatedly
with the creation operator on the cyclic vector, which physically
corresponds to the vacuum state.

Thus, even in systems involving noncommutative observables, the GNS
construction yields a Hilbert space $\mathcal{H}_{\omega}$ determined
by the state $\omega$. That is, space emerges as a structure constructed
from the observer's perspective, depending on the algebraic properties
of the observables and the choice of state. 

\section{Commutativity and the Emergence of Classical Space: A Reconsideration
via Quantum Observation}

In this appendix, we reconsider the distinction between measurable
spaces in classical theory and those in quantum theory. We argue that
the essential difference lies solely in the noncommutativity of observables:
the structure of measurable space is not fixed once and for all, but
rather emerges anew each time an observer selects a specific observable.

From this perspective, every act of observation brings forth a new
measure space---a classical (phase) space tailored to the set of
compatible observables. This leads to a context-dependent view of
space: space is not absolute, but arises from the particular measurement
context determined by the observer's choices.

However, such contextuality also raises the possibility that these
constructed measure structures may not always be consistently compatible
with one another. For further discussion of this issue, including
the BKS (Bell--Kochen--Specker) theorem and the KCBS inequality,
see the analysis by Kitano (2018) \cite{key-3}.

For a topos-theoretic treatment of contextuality, we refer the reader
to the work of Isham and D$\ddot{\mathrm{o}}$ring \cite{key-13}.

To state this viewpoint succinctly:
\begin{prop*}
Observation is the act of constructing space.
\end{prop*}
\color{black}
When taking the spectrum of a commutative observable algebra, we
assume that the selected algebra is a unital commutative
$C^*$-algebra. An unbounded self-adjoint observable is instead
treated through its spectral measure in a Hilbert-space
representation. Boundedness of the elements of a $*$-algebra alone
does not make it a $C^*$-algebra.
\color{black}

From the previous discussion, a quantum system is characterized by
a $*$-algebra $\mathcal{A}$ of observables and a state $\omega$,
where the observables $\hat{x}\in\mathcal{A}$ are self-adjoint. However,
in this abstract algebraic formulation, the quantum system is not
yet explicitly connected to a classical measure space.

In the classical case, the expectation value of an observable $\hat{x}$
is written as:

\begin{equation}
\omega(\hat{x})=\langle x\rangle_{\rho}=\int xd\mu(x)=\int x\rho(x)dx
\end{equation}
where $\rho(x)$ is a classical probability density and $\mu$ is
the associated measure.

In quantum theory, however, the expectation value takes the form:

\begin{equation}
\omega(\hat{x})=\langle\Omega_{\omega}|\pi_{\omega}(\hat{x})|\Omega_{\omega}\rangle=\mathrm{Tr}|\Omega_{\omega}\rangle\langle\Omega_{\omega}|\pi_{\omega}(\hat{x})\equiv\mathrm{Tr}\left(\hat{\rho}_{\omega}\pi_{\omega}(\hat{x})\right).
\end{equation}

Here, the object $\hat{\rho}_{\omega}=|\Omega_{\omega}\rangle\langle\Omega_{\omega}|$
is a density matrix, which serves as the quantum analogue of the classical
probability density $\rho$. It encodes all the statistical information
of the quantum state and allows us to express expectations in terms
of operator traces rather than integrals over a predefined space.

Let us now examine how objects such as $x$, the underlying set, and
$(X,\,\mathcal{B})$, the measurable space, appear within quantum
theory. This requires a somewhat formal discussion, but it can be
\color{black}expressed in terms of the spectral measure of $\hat{x}$.\color{black}

The crucial point is that this construction depends on the choice
of observable $\hat{x}$.
\color{black}
Let $E_{\hat{x}}(\Delta)$ be its spectral projection for a Borel set
$\Delta\subseteq\mathbb R$. In the state $\Omega_{\omega}$, define the
probability measure of outcomes by
\begin{equation}
\mu_{\hat{x}}(\Delta)
=\langle\Omega_{\omega}|E_{\hat{x}}(\Delta)|\Omega_{\omega}\rangle,
\qquad \Delta\in\operatorname{Borel}(\mathbb R).
\end{equation}
When the expectation exists, it is given by
\begin{equation}
\omega(\hat{x})=\int_{\mathbb R}\lambda\,d\mu_{\hat{x}}(\lambda).
\end{equation}
The underlying outcome space may be taken as the spectrum
$X(\hat{x})=\operatorname{spec}(\hat{x})$, with its Borel
$\sigma$-algebra $\mathcal B(\hat{x})$ and the measure $\mu_{\hat{x}}$
restricted to that spectrum. This gives the measurable space
\begin{equation}
\bigl(X(\hat{x}),\mathcal B(\hat{x}),\mu_{\hat{x}}\bigr).
\end{equation}
\color{black}

This reveals a profound difference between classical and quantum theories:
namely, the role of noncommutativity. In classical mechanics, all
observables commute and share a common measurable space. In quantum
theory, however, each observable (or set of commuting observables)
generates its own context-specific measure space---there is no globally
fixed underlying space independent of observational choice.

For example, in classical mechanics, the expectation values of position
$x$ and and momentum $p$ are given by:
\begin{equation}
\omega_{\rho}(\hat{x})=\int x\rho(x,p)dxdp=\int xd\mu(x,p),
\end{equation}

\begin{equation}
\omega_{\rho}(\hat{p})=\int p\rho(x,p)dxdp=\int pd\mu(x,p),
\end{equation}
where $(x,p)$ are points in phase space, distinguished by an appropriate
Borel $\sigma$-algebra $\mathcal{B}$, and where the underlying space---here,
the phase space---is denoted by $\Omega$. In this way, we obtain
a classical measure space $(\Omega,\ \mathcal{B},\ \mu)$.

\color{black}In the Koopman--von Neumann (KvN) representation\cite{key-Koopman}\cite{key-Neumann}, the classical position and momentum observables act as multiplication operators on phase-space wavefunctions and therefore commute:\color{black}
\begin{equation}
[\hat{x},\hat{p}]=0.
\end{equation}
\color{black}The auxiliary KvN operators $\hat{\pi}_x$ and $\hat{\pi}_p$ satisfy, in units where $\hbar=1$,
\begin{equation}
[\hat{x},\hat{\pi}_x]=i,\qquad [\hat{p},\hat{\pi}_p]=i.
\end{equation}
Thus the algebra of classical observables acting by multiplication is commutative, although the larger KvN operator algebra contains noncommuting auxiliary operators.\color{black}

\color{black}
The expectations of these commuting KvN multiplication
observables are given by the phase-space integrals above.
They are therefore described on the common classical measurable
space $(\Omega,\mathcal B,\mu)$.
\color{black}

\color{black}
For a unital commutative $C^*$-algebra $\mathcal C_M$, the
Gelfand--Naimark theorem identifies $\mathcal C_M$ with the algebra
of continuous complex-valued functions on its spectrum
$\operatorname{Spec}(\mathcal C_M)$. Here this theorem relates the
selected commutative algebra to its spectrum.
\color{black}

From the perspective of (phase) space, quantum theory inverts the
classical logic: it begins not with a pre-given space, but with a
$*$-algebra $\mathcal{A}$ of operators (i.e., q-numbers) and a state
$\omega$ or density operator $\hat{\rho}$ that defines expectation
values. In other words, quantum theory is specified by a pair $(\mathcal{A},\omega$),
where structure and dynamics are encoded algebraically.

Given this, the observer selects a self-adjoint observable $\hat{a}\in\mathcal{A}$
(i.e., $\hat{a}=\hat{a}^{\dagger})$ to measure. Each such choice
induces a corresponding measure space:
\begin{equation}
(X(\hat{a}),B(\hat{a}),\mu(\hat{a})),
\end{equation}
which may be interpreted as a classical world emergent from the context
defined by $\hat{a}$.

This viewpoint---that the structure of space emerges from the observer's
act of measurement---resonates with Rovelli's relational quantum
mechanics \cite{key-9,key-10,key-16}, where space is not treated
as a fixed background but as a web of relations defined through interactions.
In this sense, our approach represents a serious reevaluation of the
quantum worldview, combining the Heisenberg picture with the Born
rule and grounding spatial structure in operational choices.

This leads us naturally to further questions:
\begin{problem*}
What kinds of sets $\{\hat{a}_{i}\}_{i=1}^{N}$ of observables can
be simultaneously measured in quantum theory? How do these sets transform
under time evolution?
\end{problem*}
The structure of such commutative subalgebras within a noncommutative
algebra is a deep subject in quantum foundations. The pioneering results
of Kochen and Specker \cite{key-15} showed that no global assignment
of values to all observables is possible in general, due to contextuality.
For a detailed philosophical and information-theoretic discussion,
we refer the reader to Bub's analysis \cite{key-4}, where the logical
structure and classical reality conditions of quantum theory are thoroughly
examined.


\begin{thebibliography}{10}
\bibitem{key-1}Kato, Fumiharu. ``What is Space?'', Gendai Shiso (Japanese),
Seidosha, January 2024.

\bibitem{key-5}Jost, J$\ddot{\mathrm{u}}$rgen. ``Mathematical concepts'',
Springer, 2015.

\bibitem{key-2}So Katagiri, ``On the Dynamical View of Mathematics
and the Quantum-Theoretical World Picture'', in Mathematical Physics
Research II, S$\overline{\mathrm{u}}$ri-Butsuri Publishing Association,
Tokyo, 2023. ISBN:978-4-9911190-2-6 (in Japanese)

\bibitem{key-12}Bratteli, Ola, and Derek William Robinson. ``Operator
algebras and quantum statistical mechanics: Volume 1: $C^{*}$-and
$W^{*}$-Algebras.'', Symmetry Groups. Decomposition of States. Springer
Science \& Business Media, 2012.

\bibitem{key-7}Connes, Alain. ``Noncommutative geometry'', Academic
Press, 1994.

\bibitem{key-17}Gracia-Bond$\mathrm{\acute{i}}$a, Jos$\acute{\mathrm{e}}$
M., Joseph C. V\'{a}rilly, and H$\acute{\mathrm{e}}$ctor Figueroa. ``Elements
of noncommutative geometry'', Springer Science \& Business Media,
2013.

\bibitem{key-3}Yuichiro Kitajima, ``Contextuality in Algebraic Quantum
Theory'', Journal of the Japan Association for Philosophy of Science
(Japanese), Vol. 45, Nos. 1 \& 2, 2018.

\bibitem{key-13}C. J. Isham \& A. D$\ddot{\mathrm{o}}$ring,``What
is a Thing? Topos Theory in the Foundations of Physics'', in New Structures
for Physics, Springer (2011)

\bibitem{key-9}Rovelli, Carlo. ``Quantum gravity'', Cambridge university
press, 2004.

\bibitem{key-10}Rovelli, Carlo. ``The order of time``, Penguin, 2019.

\bibitem{key-16}Rovelli, Carlo. ``Relational quantum mechanics'',
International journal of theoretical physics 35 (1996): 1637-1678.

\bibitem{key-20}Everett III, Hugh. ```` Relative state'' formulation
of quantum mechanics'', Reviews of modern physics 29.3 (1957): 454.

\bibitem{key-18}Katagiri, So. ``Measurement theory in classical mechanics'',
Progress of Theoretical and Experimental Physics 2020.6 (2020): 063A02.

\bibitem{Born-1}Max, Born  ``suggestion for unifying quantum theory and relativityProc.'', R. Soc. Lond. (1938)A165291-303.

\bibitem{Boyle-1}Boyle, Michael, Robert Owen, and Harald P. Pfeiffer. ``Geometric approach to the precession of compact binaries.'', Physical Review D - Particles, Fields, Gravitation, and Cosmology 84.12 (2011): 124011.

\bibitem{Glikman-1}Kowalski-Glikman, Jerzy. ``Living in curved momentum space.'' International Journal of Modern Physics A 28.12 (2013): 1330014.

\bibitem{Magid-1} Majid, Shahn. ``Algebraic approach to quantum gravity I : relative realism.'',  In James Ladyman, Stuart Presnell, Gordon McCabe, Michal Eckstein \& Sebastian J. Szybka, Road to reality with Roger Penrose. Krakow: Copernicus Center Press(2015).

\bibitem{Lieb-1}Lieb, Elliott H., and Jakob Yngvason. "The physics and mathematics of the second law of thermodynamics.",
Physics Reports 310.1 (1999): 1-96.

\bibitem{key-dec}Zurek, Wojciech Hubert. "Decoherence, einselection, and the quantum origins of the classical." Reviews of modern physics 75.3 (2003): 715.

\bibitem{key-ozawa}Ozawa, Masanao. "Universally valid reformulation of the Heisenberg uncertainty principle on noise and disturbance in measurement." Physical Review A 67.4 (2003): 042105.

\bibitem{key-POVM}Davies, E. Brian, and John T. Lewis. "An operational approach to quantum probability." Communications in Mathematical Physics 17.3 (1970): 239-260.

\bibitem{key-Ojima}Ojima, Izumi, and Mitsuharu Takeori. "How to observe quantum fields and recover them from observational data? Takesaki duality as a micro-macro duality." Open Systems \& Information Dynamics 14.03 (2007): 307-318.

\bibitem{key-Ojima2}Ojima, Izumi. "Micro-macro duality and emergence of macroscopic levels." Quantum Bio-Informatics: From Quantum Information to Bio-Informatics. 2008. 217-228.

\bibitem{key-Saigo}Saigo, Hayato. "Category algebras and states on categories." Symmetry 13.7 (2021): 1172.

\bibitem{key-Saigo2}Saigo, Hayato. "Quantum Fields as category algebras." Symmetry 13.9 (2021): 1727.

\bibitem{key-Koopman}B. O. Koopman, ``Hamiltonian Systems and Transformation in Hilbert Space'' Proceedings of the National Academy of Sciences 17 (5), 315-318 (1931). DOI: 10.1073/pnas.17.5.315.

\bibitem{key-Neumann}J. von Neumann, ``Zur Operatorenmethode in der klassischen Mechanik'' Annals of Mathematics 33 (3), 587-642 (1932). DOI: 10.2307/1968537.

\bibitem{key-15}Kochen, Simon, and Ernst P. Specker. ``The problem
of hidden variables in quantum mechanics'', Ernst Specker Selecta
(1990): 235-263.

\bibitem{key-4}Bub, Jeffrey. ``Interpreting the quantum world'',
Cambridge University Press, 1999.


\end{thebibliography}
\end{document}